\documentclass[pre, showpacs, aps, superscriptaddress]{revtex4}
\usepackage{amsfonts}
\usepackage{epsfig}
\newcommand{\pd}[2]{\frac{\partial #1}{\partial #2}}
\newcommand{\V}[1]{{\bf #1}}
\renewcommand{\k}{{\bf k}}
\renewcommand{\Im}{{\rm Im\ }}
\newcommand{\w}{\omega}

\begin{document}
\title{Dimensional Analysis and Weak Turbulence}
\author{Colm Connaughton}
\affiliation{Mathematics Institute, University of Warwick,
Coventry CV4 7AL, U.K.}
\author{Sergey Nazarenko}
\affiliation{Mathematics Institute, University of Warwick,
Coventry CV4 7AL, U.K.}
\author{Alan C. Newell}
\affiliation{Mathematics Institute, University of Warwick,
Coventry CV4 7AL, U.K.} \affiliation{Department of Mathematics,
University of Arizona, Tucson, AZ 85721, U.S.A.}

\begin{abstract}
In the study of weakly turbulent wave systems possessing
incomplete self-similarity it is possible to use dimensional
arguments to derive the scaling exponents of the
Kolmogorov-Zakharov spectra, provided the order of the resonant
wave interactions responsible for nonlinear energy transfer is
known. Furthermore one can easily derive conditions for the
breakdown of the weak turbulence approximation. It is found that
for incompletely self-similar systems dominated by three wave
interactions, the weak turbulence approximation usually cannot
break down at small scales. It follows that such systems cannot
exhibit small scale intermittency. For systems dominated by four
wave interactions, the incomplete self-similarity property implies
that the scaling of the interaction coefficient depends only on
the physical dimension of the system. These results are used to
build a complete picture of the scaling properties of the surface
wave problem where both gravity and surface tension play a role.
We argue that, for large values of the energy flux, there should
be two weakly turbulent scaling regions matched together via a
region of strongly nonlinear turbulence.
\end{abstract}
\pacs{04.30.Nk, 92.10.Hm, 92.10.Cg}

\maketitle

\section{Introduction}
\label{sec-intro}

The time evolution of the average spectral wave-action density,
$n_\k$,  of an ensemble of weakly interacting dispersive waves is
governed by the so-called kinetic equation. For a system with
dispersion law $\omega_\k$, dominated by 3-wave interactions with
interaction coefficient $L_{\k\k_1\k_2}$, the kinetic equation is
:
\begin{equation}
\label{eq-3WKE} \pd{n_\k}{t} = 4\pi \int
\left|L_{\k\k_1\k_2}\right|^2 n_\k n_{\k_1} n_{\k_2}{\mathcal
F}_3\left[n\right]\delta(\k-\k_1-\k_2)\ d\k_1d\k_2
\end{equation}
where
\begin{eqnarray}
\nonumber {\mathcal F}_3\left[n\right]&=& \left(\frac{1}{n_\k}-\frac{1}{n_{\k_1}}-\frac{1}{n_{\k_2}}\right)\delta(\omega_\k-\omega_{\k_1}-\omega_{\k_2})\\
&+&\left(\frac{1}{n_\k}-\frac{1}{n_{\k_1}}+\frac{1}{n_{\k_2}}\right)\delta(\omega_{\k_1}-\omega_{\k}-\omega_{\k_2})\\
\nonumber&+&\left(\frac{1}{n_\k}+\frac{1}{n_{\k_1}}-\frac{1}{n_{\k_2}}\right)\delta(\omega_{\k_2}-\omega_{\k}-\omega_{\k_1}).
\end{eqnarray}
More generally, if the dominant nonlinear interaction is $N$-wave,
the {\it schematic} form of the kinetic equation is
\begin{equation}
\label{eq-NWKE} \pd{n_\k}{t} \sim \int  L^2_N\,
n_\k^{N-1}\,\delta(\omega_\k)\, \delta(\k)\, (d\k)^{N-1}.
\end{equation}
Let us consider homogeneous, isotropic systems in physical
dimension $d$. Let us further assume scale invariance with
$\omega_\k$ and $L_N$ homogeneous functions of their arguments.
Denote the degrees of homogeneity of the dispersion, $\omega_\k$,
and the $N$-wave interaction coefficient, $L_N$, by $\alpha$ and
$\gamma_N$ respectively. Under these assumptions, the kinetic
equation possesses exact stationary solutions, found originally by
Zakharov in the early 70's, which carry constant fluxes of
conserved quantities, such as energy or wave-action. These
solutions are called Kolmogorov-Zakharov (KZ) spectra.

The 3-wave kinetic equation admits a single KZ spectrum carrying a
constant flux, $P$, of energy :
\begin{equation}
\label{eq-3WKZ} n_\k = c^{(3)} P^\frac{1}{2} k^{-(\gamma_3+d)}.
\end{equation}
The 4-wave kinetic equation conserves wave-action in addition to
energy and thus admits a pair of KZ spectra, one carrying an
energy flux, $P$, the other carrying a wave-action flux, $Q$. They
are
\begin{eqnarray}
\label{eq-4W-PKZ}n_\k &=& c^{(4)}_1 P^\frac{1}{3} k^{-\frac{1}{3}(2\gamma_4+3d)},\\
\label{eq-4W-QKZ}n_\k &=& c^{(4)}_2 Q^\frac{1}{3}
k^{-\frac{1}{3}(2\gamma_4+3d-\alpha)}.
\end{eqnarray}
The dimensional constants, $C^{(N)}$, can be explicitly
calculated.

Suppose we know that the physical system under consideration
depends on only one dimensional constant. Such a wave system is
said to possess  {\it incomplete self-similarity} (ISS). Zakharov,
Lvov and Falkovich have pointed out \cite[chap. 3]{zakharov} that
for such systems the scaling of the KZ spectra can be obtained
from a dimensional argument. The dimensional argument for ISS
systems uses only the scaling of the dispersion relation and not
the scaling of the interaction coefficients required for the more
general scale invariant systems considered above. This fact has
not been fully appreciated and is rarely used, despite the fact
that most of the known wave turbulence systems are ISS, as will be
shown in this paper.

\noindent The dimensional analysis determines the scaling of the
interaction coefficients for systems possessing ISS, a point that
has been mostly overlooked before. The fact that the scaling
exponent of the interaction coefficient is not an independent
quantity may have some consequences for the practical
applicability of some theoretical results on weak turbulence,
where the scaling exponents of the interaction coefficients are
regarded as arbitrary. In particular, Biven, Nazarenko and
Newell\cite{biven2001,newell2001} have recently pointed out that
the weak turbulence approximation is almost never uniformly valid
in $k$ but rather breaks down either at large or small scales. The
breakdown of weak turbulence at small scales is presumed to signal
the onset of small scale intermittency. It is possible to use a
simple dimensional argument to recover the criteria obtained in
\cite{biven2001} in the ISS case. One finds that the condition for
breakdown at small scales is inconsistent for three wave systems.
As a result one would not expect such systems to exhibit small
scale intermittency.

\noindent The goal of this paper is to use the ISS dimensional
argument, for the first time in some examples, to derive the KZ
spectra and the scaling of the interaction coefficients for a
large number of commonly considered applications of weak
turbulence. We then use our results to discuss the uniformity of
the weak turbulence approximation in $\k$ for these physical
systems. In the final section we consider the water wave system in
more detail. It is shown that by considering the effect of both
the gravity dominated and surface tension dominated parts of the
spectrum together, one can build a consistent picture of energy
transfer in the system, even when the flux is sufficiently large
to cause breakdown of the weak turbulence approximation.

\section{Dimensional Derivation of Kolmogorov-Zakharov Spectra}
\label{sec-KS} Before we begin, let us clarify a point of
notation. We deal with isotropic systems. Physical quantities such
as spectral wave-action density, $n_\k$, or spectral energy
density, $E_\k=\w_\k n_\k$, only depend on the modulus, $k$, of
the wave-vector, $\k$. It is often convenient to integrate over
angles in $\k$-space. We need to make a distinction between a
spectral quantity which has been averaged over angle and one which
has not. To do this, we use a regular type argument to denote a
quantity which has been integrated over angles, as in $n_k$, and a
bold type argument to denote one which has not, as in $n_\k$. The
two are easily related. Consider for example, the wave-action
density:
\begin{eqnarray}
\nonumber \int n_k\, dk &=& \int n_\k\,d\k\\
\Rightarrow n_k &=& \Omega_D\,n_\k k^{D-1},
\end{eqnarray}
where $\Omega_D$ is the solid angle coming from the integration
over angles in $D$-dimensional wave-vector space. We shall use $C$
to denote a generic dimensionless constant whose value cannot be
determined from dimensional arguments.

\subsection{Constant Energy Flux Spectra}
\label{subsec-energySpectra}
Suppose we have a wave system characterised by a single additional
dimensional parameter, $\lambda$, which appears in the dispersion relation
in the form
\begin{equation}
\label{eq-disp} \omega_\k=\lambda k^\alpha .
\end{equation}
It is convenient to set the density of the medium to 1. Our unit
of mass then has dimension ${\rm L}^3$ and energy has dimension
${\rm L}^5{\rm T}^{-2}$. We suppose that the $d$-dimensional
energy density, ${\mathcal E}$, is finite in physical space. For
example, $d=2$ for water waves while $d=3$ for acoustic waves. We
denote the dimension of the Fourier transform used to go to a
spectral description of the theory by $D$. Usually $D=d$ but not
always (see, for example, section \ref{subsec-5wave}). The
spectral energy density, $E_k$, is defined by
\begin{equation}
{\mathcal E} = \int E_\k\,d\k = \int_0^\infty  E_k\,dk.
\end{equation}
$E_k$ clearly has dimension ${\rm L}^{6-d}{\rm T}^{-2}$. The
energy flux, $P$, has dimension ${\rm L}^{5-d}{\rm T}^{-3}$ and
$\lambda$  has dimension ${\rm L}^{\alpha}{\rm T}^{-1}$. \noindent
Let us now consider the constant energy flux spectrum for this
system. For 3-wave processes, the energy flux is proportional to
the square of the spectral energy density so we can write
\begin{equation}
\label{eq-3waveE1} E_k = C\sqrt{P}\lambda^{\rm X} k^{\rm Y},
\end{equation}
where  $C$ is a dimensionless constant and the exponents ${\rm X}$ and
${\rm Y}$ are to be determined by dimensional analysis. This yields
\begin{equation}
\label{eq-exp1}
{\rm X}=\frac{1}{2},\hspace{0.25in} {\rm Y} = \frac{1}{2}\left(d + \alpha -7\right).
\end{equation}
This argument, used by Kraichnan \cite{kraichnan65} in the context
of Alfv\'en waves, can be generalised to $N$-wave systems. In a
system dominated by $N$ wave processes, the energy flux is
proportional to the $N-1^{\rm th}$ power of the spectral energy
density and a similar argument yields the scaling law
\begin{equation}
\label{eq-NwaveE} E_k = CP^\frac{1}{N-1}\lambda^{\rm X} k^{\rm Y},
\end{equation}
with ${\rm X}$ and ${\rm Y}$ given by
\begin{eqnarray}
\label{eq-exp2}
\nonumber & &{\rm X}=\frac{2N-5}{N-1},\\
& &{\rm Y} = (2\alpha + d -6) +\frac{5-3\alpha -d}{N-1}.
\end{eqnarray}
Associated with each constant energy flux spectrum, we have a
particle number (wave action) spectrum, $n_\k$. One can be
obtained from the other via the relation
\begin{equation}
\label{eq-energyVsParticle} \int_0^\infty E_k\,dk =
\Omega_D\int_0^\infty\omega_\k\,n_\k\,k^{D-1}\,dk,
\end{equation}
where $\Omega_D$ is the $D$-dimensional solid angle. The resulting
scaling law for $n_\k$ for an  $N$ wave system is
\begin{equation}
\label{eq-NwaveN} n_\k = CP^\frac{1}{N-1}\lambda^{\rm X}k^{\rm Y},
\end{equation}
where
\begin{eqnarray}
\label{eq-exp3}
\nonumber & &{\rm X}=\frac{N-4}{N-1},\\
& &{\rm Y} = (\alpha + d -D  -5) +\frac{5-3\alpha -d}{N-1}.
\end{eqnarray}

\subsection{Constant Particle Flux Spectra}
\label{subsec-particleSpectra} In the case of a system with 4-wave
interactions, the total particle number, $N=\int n_k\,dk$ is also
a conserved quantity. As a result, there can also exist a
constant-flux spectrum carrying a flux of particles rather than
energy. Such behaviour is associated with a continuity equation of
the form
\begin{equation}
\label{eq-flux2} \pd{n_k}{t} +  \pd{Q_k}{k}=0,
\end{equation}
where $Q_k$ is the particle flux. One can perform the same
dimensional analysis for this spectrum, bearing in mind that
dimensionally, $P=\omega_\k Q$. One obtains the following spectrum
describing a constant flux of particles:
\begin{equation}
n_k = CQ^\frac{1}{3} \lambda^\frac{1}{3}k^{-\frac{1}{3}\left(
-2d-\alpha+13\right)},
\end{equation}
or
\begin{equation}
n_\k = CQ^\frac{1}{3} \lambda^\frac{1}{3}k^{-\frac{1}{3}\left(
3D-2d-\alpha+10\right)}.
\end{equation}

\subsection{Scaling of the Interaction Coefficients}
\label{subsec-interaction}
In the regime where the system is scale invariant, the
nonlinear interaction coefficients, $V_{ijk}$ (3-wave) and $T_{ijkl}$
(4-wave) often possess nontrivial scaling properties. For the 3-wave case
we have :
\begin{equation}
\label{eq-V} V_{h \V{k}h\V{k}_1h \V{k}_2} = h^\beta
V_{\V{k}\V{k}_1\V{k}_2},
\end{equation}
and for the 4-wave case :
\begin{equation}
\label{eq-T} T_{h \V{k}h\V{k}_1h\V{k}_2h\V{k}_3}= h^\gamma
T_{\V{k}\V{k}_1\V{k}_2\V{k}_3}.
\end{equation}
In fact $\beta$ and $\gamma$ cannot be arbitrary. They may be
determined from dimensional analysis of the dynamical equations.
Schematically, the dynamical equations for an $N$ wave system look
like
\begin{equation}
\frac{\partial a_\k}{\partial t} + i\omega_\k a_\k = \int L_{N}\,
a_\k^{N-1}\,\delta(\k)\,(d\k)^{N-1}.
\end{equation}
Recalling that dimensionally, $\left[\k\right]={\rm L}^{-D}$ and
$\left[\delta(\k)\right]={\rm L}^{D}$, we see that:
\begin{displaymath}
\left[L_{N}\right] = \frac{\left[\omega_\k\right]}{\left[a_\k
\right]^{N-2}\left[{\rm L}^{-D}\right]^{N-2}}.
\end{displaymath}
Determine the dimension of $a_\k$ as follows,
\begin{displaymath}
<a_\k a^*_{\k^\prime}> = \delta(\k-\k^\prime)\,n_\k.
\end{displaymath}
So
\begin{displaymath}
\left[a_\k\right]^2 =
\left[\delta(\k-\k^\prime)\right]\frac{\left[E_k\right]}{\left[\omega_\k\right]
{\rm L}^{1-D}} = \frac{{\rm L}^{6-d}{\rm T}^{-2}}{{\rm T}^{-1}
{\rm L}^{1-2D}}.
\end{displaymath}
This results in the following expression for the dimension of the interaction
coefficient
\begin{equation}
\label{eq-interactionScaling} [L_N] = {\rm
T}^{\frac{1}{2}(N-4)}{\rm L}^{\frac{1}{2}(N-2)(d-5)},
\end{equation}
and dimensional analysis then yields,
\begin{equation}
\label{eq-interaction formula} L_{N \,{\k_1\cdots\k_N}}=
\lambda^{\frac{1}{2}(4-N)}k^{\gamma_N}\,f_{\k_1\cdots\k_N},
\end{equation}
where
\begin{equation}
\gamma_N = -\frac{1}{2}\left\{(N-2)(d-5)+(N-4)\alpha\right\}.
\end{equation}
Here $f_{\k_1\cdots\k_N}$ is a dimensionless function of
$\k_1\cdots\k_N$. Interestingly, for 4-wave systems, $N=4$, the
scaling of the interaction coefficients depends only on the
dimension, $d$, of the system and is independent of any
dimensional parameter, including $\lambda$. We see that {\it all}
incompletely self-similar 4-wave systems exhibit the same scaling
behaviour of their interaction coefficients,
\begin{equation}
\gamma_4 = 5-d.
\end{equation}

\noindent Applying our analysis to the 3-wave case yields
$L_{3\,\k_1\k_2\k_3} \sim k^{\gamma_3}$, where
\begin{equation}
\gamma_3 = \frac{1}{2}\left(5+\alpha-d\right).
\end{equation}
The scaling in this case  depends on the dispersion index,
$\alpha$, but we see that $\gamma_3$ and $\alpha$ are not
independent quantities. This fact, while obvious from this point
of view, is possibly not fully appreciated. We shall see that the
class of incompletely self-similar systems for which this analysis
is valid  includes most of the common physical applications of
weak turbulence.

\section{Breakdown of the Weak Turbulence Approximation}
\label{sec-breakdown} By analysing the scaling behaviour of the
kinetic equations describing nonlinear energy transfer in weak
turbulence, Biven, Nazarenko and Newell \cite{biven2001} have,
under certain assumptions, given a set of criteria for the
breakdown of the weak turbulence approximation. These assumptions
are that the turbulent transfer is sufficiently local that after
using homogeneity properties to remove the k-dependence of the
collision integral and the other integrals arising in the
expression for the frequency renormalisation, the remaining
integrals converge. We discuss in the conclusion and in
\cite{biven2002} how this may not always be the case when the
coefficient of long-wave short-wave interaction is too strong.

In this section we shall adopt the commonly used notation
$\gamma_3=\beta$ and $\gamma_4=\gamma$. For three wave systems,
breakdown occurs at small scales for $\beta-2\alpha>0$ and at
large scales for $\beta-2\alpha<0$. For four wave systems
breakdown occurs at small scales for $\gamma-3\alpha>0$ in the
presence of a pure energy flux and for $\gamma-2\alpha>0$ in the
case of a pure particle flux. The breakdown at large scales can be
masked by the large scale forcing but breakdown at small scales,
in the absence of dissipation, is taken to signal the onset of
small scale intermittency. \noindent In the case of ISS systems we
can construct a characteristic scale, $k_{\rm NL}$, from the flux
and the parameter $\lambda$. From our previous discussion of
dimensions, we see that the quantity,
$(P/\lambda^3)^{1/5-d-3\alpha}$ has the dimension of a length.
Thus in the case of a finite energy flux, we define
\begin{equation}
\label{eq-breakdown} k_{\rm NL} =
\left(\frac{P}{\lambda^3}\right)^{-\frac{1}{5-d-3\alpha}}.
\end{equation}
Likewise, in the case of a finite particle number flux, $Q$, in a four wave
system we define
\begin{equation}
\label{eq-breakdownQ} k_{\rm NL} =
\left(\frac{Q}{\lambda^3}\right)^{-\frac{1}{5-d-2\alpha}}.
\end{equation}
For small fluxes, $P\to0$, we see that the breakdown occurs at
small scales for $5-d-3\alpha >0$ for finite $P$ and for
$5-d-2\alpha>0$ for finite $Q$. Upon substitution of
(\ref{eq-interactionScaling}) into these expressions, we recover
the criteria of \cite{biven2001} in terms of the scaling exponents
of the interaction coefficients. \noindent It is interesting to
note that for finite energy flux, the breakdown criterion is
$\alpha < 2/3$ in 3 dimensions and $\alpha<1$ in 2 dimensions.
However it is known from the work of Krasitskii
\cite{krasitskii91} that for $\alpha<1$, 3-wave terms in the
interaction Hamiltonian are nonresonant and can be removed by an
appropriate change of canonical variables to give an effective
description in terms of four wave interactions. Thus the small
scale breakdown criterion can never be realised for three wave
systems in two or three dimensions. This means that a significant
number of physical systems cannot be hoped to exhibit
intermittency at small scales as discussed in the following
section. Conversely, these systems can always exhibit
intermittency at large scales, provided the forcing is
sufficiently strong, without affecting the validity of the weak
turbulence approximation at small scales. At this point, it is
worth mentioning that the case $\alpha=1$ is borderline in two
dimensions. Such three wave systems, 2-d sound being an example,
are known to be rather special and must be carefully treated
separately.

\section{Examples of 3-Wave Systems}
\label{subsec-3wave}
\subsection*{Sound and magnetic sound in 3 dimensions}
Acoustic turbulence \cite{zakharov70,newell71,lvov97} corresponds
to the almost linear dispersion $\omega_\k\approx ck$, where $c$
is the sound speed or magnetic sound speed, so that $\alpha=1$ and
$d=3$. We thus obtain the following pair of spectra for the energy
and wave action
\begin{equation}
\label{eq-sound3D} E_k = C\sqrt{Pc}k^{-\frac{3}{2}},
\hspace{0.5in} n_\k=C^\prime\sqrt{\frac{P}{c}}k^{-\frac{9}{2}}.
\end{equation}
These are the original spectra obtained by Zakharov and Sagdeev.
According to our analysis, this spectrum remains uniformly valid
at small scales.
\subsection*{3-D Alfv\'{e}n waves}
3-D Alfv\'{e}n wave turbulence was originally considered by
Iroshnikov \cite{iroshnikov65} and Kraichnan \cite{kraichnan65} in
the 60's . Such waves are also weakly dispersive and from the
point of view of the dimensional analysis are identical to
acoustic waves discussed above. The resulting $-\frac{3}{2}$ and
$-\frac{9}{2}$ spectra are not actually realised in real plasmas
because the true Alfv\'{e}n wave turbulence is anisotropic.
\subsection*{Quasi 2-D Alfv\'{e}n waves}
In reality, the Alfv\'{e}n turbulence is strongly anisotropic and is
described by  quasi-2D rather than 3D spectra.
For this system we have again $\alpha=1$ but $d=2$. This yields the
stationary spectra
\begin{equation}
\label{eq-alfven2D} E_k = C\sqrt{Pc}k^{-2}, \hspace{0.5in}
n_\k=C^\prime\sqrt{\frac{P}{c}}k^{-4},
\end{equation}
which are the spectra obtained using a  dimensional analysis by Ng and
 Bhatachargee \cite{ng96} and analytically derived by Galtier,
Nazarenko and Newell \cite{galtier2000}. As mentioned already, the
case $\alpha=1$ is borderline in 2-d so that our argument
concerning the breakdown of the weak turbulence approximation
remains inconclusive. In fact, unlike typical three wave systems,
this system does exhibit breakdown at small scales as shown in
\cite{galtier2000}.

\subsection*{Capillary waves on deep water}
$\lambda=\sqrt{\sigma}$, where $\sigma$ is the coefficient of surface tension.
In this case, $\alpha=\frac{3}{2}$, $d=2$ and the Kolmogorov spectrum is
\begin{equation}
\label{eq-cap2D} E_k =
C\sqrt{P}\sigma^\frac{1}{4}k^{-\frac{7}{4}}, \hspace{0.5in}
n_\k=C^\prime\sqrt{P}\sigma^{-\frac{1}{4}}k^{-\frac{17}{4}}.
\end{equation}
This spectrum was first derived by Zakharov and Filonenko
\cite{zakharov67}. There is no small scale intermittency in this
system.
\section{Examples of 4-Wave Systems}
\label{subsec-4wave}
\subsection*{Gravity waves on deep water}
For this system, $\alpha=\frac{1}{2}$, $d=2$ and $\lambda=\sqrt{g}$, where
$g$ is the gravitational constant. The Kolmogorov spectrum
corresponding to a constant flux of energy is then
\begin{equation}
\label{eq-grav} E_k = CP^\frac{1}{3}g^\frac{1}{2}k^{-\frac{5}{2}},
\hspace{0.5in} n_\k=C^\prime P^\frac{1}{3}k^{-4}.
\end{equation}
There is also a second spectrum corresponding to a constant flux of wave action,
\begin{equation}
\label{eq-grav2} n_\k =
CQ^\frac{1}{3}g^{\frac{1}{6}}k^{-\frac{23}{6}}.
\end{equation}
These spectra were obtained by Zakharov and Filonenko
\cite{zakharov66}. In this case, the energy spectrum breaks down
at small scales.

This is one of the cases where certain integrals appearing in the
frequency renormalisation series diverge on the K-Z spectrum. The
problem is that the interaction coefficient between the high k
mode and long wave partners in its resonant quartet is too strong.
This leads to a modification of the breakdown criterion and means
that the breakdown can occur for values of k less than that value
calculated when local interactions dominate.

\subsection*{Langmuir waves in isotropic plasmas, spin waves}
Langmuir waves are described by the dispersion relation
\begin{equation}
\omega^2_\k = \omega^2_p\left(1+3r_D^2k^2 \right),
\end{equation}
where $\omega_p$ and $r_D$ are the plasma frequency and Debye
length respectively. Magnetic spin waves in solids also obey a
dispersion relation of this type but the physical meaning of the
dimensional parameters is different. \cite{lvov87},
\cite{zakharov} For long Langmuir waves we can Taylor expand
$\omega_\k$ as $\omega_\k = \omega_p +
\frac{3}{2}\omega_pr_D^2k^2$. The constant factor $\omega_p$
cancels out of both sides of the 4-wave resonance condition so
that the effective dispersion is $\omega \sim k^2$. Thus taking
$\lambda=\w_p r_D^2$, $\alpha=2$, $d=3$ we obtain the energy
spectrum
\begin{equation}
\label{eq-langmuir} E_k =
C\,P^\frac{1}{3}\,(\omega_pr_D^2)\,k^{-\frac{1}{3}},
\hspace{0.5in} n_\k=C^\prime\, P^\frac{1}{3}\,k^{-\frac{13}{3}}.
\end{equation}
Using equation (\ref{eq-breakdown}), $k_{\rm NL}\sim (P/\w_p^3
r_D^6)^{1/4}\to 0$ as $P\to 0$ so this spectrum should break down
at large scales. The second spectrum carrying the wave action flux
is
\begin{equation}
\label{eq-langmuir2} n_\k =
C\,Q^\frac{1}{3}\,(\omega_pr_D^2)^{\frac{1}{3}}\,k^{-\frac{11}{3}},
\end{equation}
which should also break down at large scales since $k_{\rm NL}\sim
(Q/\w_p^3 r_D^6)^{1/2} \to 0$ as $Q\to 0$ from equation
(\ref{eq-breakdownQ}). These spectra were originally derived by
Zakharov \cite{zakharov72} (1972).

\section{A 5-Wave Example}
\label{subsec-5wave}
One-dimensional gravity waves were considered by Dyachenko et al
\cite{dyachenko95}. They found that the 4-wave interaction coefficient
is identically zero on the resonant manifolds so that the nonlinear
exchange of energy in the system is, in fact, due to 5-wave interactions.
This system is an example of a case where the dimension of the physical
energy density differs from the dimension of the Fourier space. In this case,
$d=2$ but $D=1$. For this system, $\lambda = \sqrt{g}$ and $\alpha =\frac{1}{2}$.
Applying (\ref{eq-NwaveN}) and (\ref{eq-NwaveE}) we obtain the finite energy
flux spectrum
\begin{eqnarray}
\label{eq-1dww}
E_k &=& C P^\frac{1}{4}g^\frac{5}{8}k^\frac{-21}{8},\\
\nonumber n_\k &=& C^\prime
(P\sqrt{g})^\frac{1}{4}k^\frac{-25}{8},
\end{eqnarray}
as found in \cite{dyachenko95}.  From equation
(\ref{eq-breakdown}), $k_{\rm NL} \sim P^{-2/3}g$ goes to large
$\k$ for small $P$. Therefore this spectrum should break down at
small scales.

\section{Matching the Gravity and Capillary Wave Spectra}
\label{sec-matching}
\begin{figure}
\begin{center}
\epsfig{file=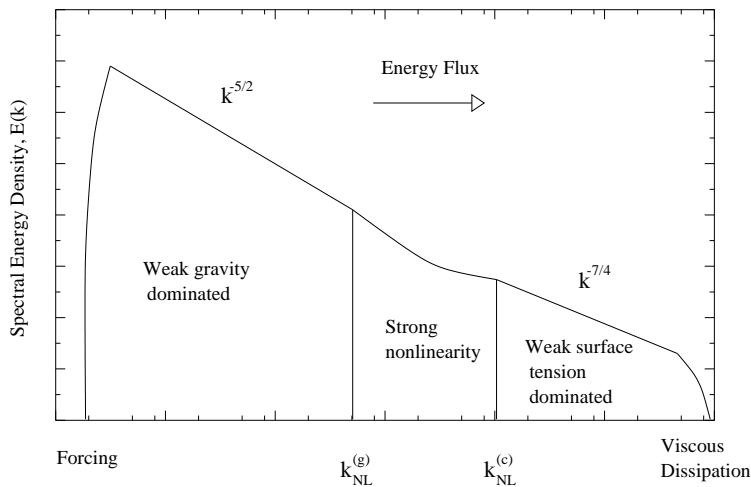, height=2.5in, angle=0}
\end{center}
\caption{\label{fig-window}Schematic representation of the different scaling regimes in the surface wave problem. The width of the window of fully nonlinear turbulence depends on the energy flux input at the forcing scale.}
\end{figure}
Let us now consider the complete surface wave problem including
both gravity and surface tension effects. At large scales the
system is entirely gravity dominated. We assume that the forcing
is at large scales only. At small scales the system is entirely
surface tension dominated down to the viscous scale where the wave
energy is finally dissipated. The characteristic scale, $k_0$,
where surface tension and gravity are comparable can be estimated
from the dispersion relation,
\begin{equation}
\label{eq-dispersion} \omega(k) = \sqrt{gk+\sigma k^3}.
\end{equation}
The gravity and surface tension effects are of comparable order when
\begin{equation}
\label{eq-k_0}k \approx k_0 = \sqrt{\frac{g}{\sigma}}.
\end{equation}
We expect that for $k<<k_0$ the system is well described by the gravity
wave spectrum (\ref{eq-grav}), for $k>>k_0$ the system is well described
by the capillary wave spectrum, (\ref{eq-cap2D}). In between there is
a non-scale invariant cross-over regime.
Let us consider the question of whether the weak turbulence approximation
 remains consistent through this cross-over regime. In order for the
turbulence to remain weak as we approach $k_0$ from the left, the gravity
wave spectrum must remain valid at least to the scale $k_0$ where surface
tension effects can start carrying the flux. Thus we require
\begin{equation}
k_{\rm NL}^{(g)} > k_0
\end{equation}
where $k_{\rm NL}^{(g)}$ is the breakdown scale for pure gravity
waves which we calculate from (\ref{eq-breakdown}) :
\begin{equation}
\label{eq-k_NL_gravity}
k_{\rm NL}^{(g)} \approx P^{-\frac{2}{3}}g.
\end{equation}
Using expressions (\ref{eq-k_NL_gravity}) and (\ref{eq-k_0}) this gives us
a condition on the flux,
\begin{equation}
\label{eq-flux}
P<(g\sigma)^\frac{3}{4}
\end{equation}
In order for the
turbulence to remain weak as we approach $k_0$ from the right, the capillary
wave spectrum should be valid by the time we reach scale $k_0$ so that it
can connect to the gravity wave spectrum. Thus we require
\begin{equation}
k_{\rm NL}^{(c)} < k_0
\end{equation}
where $k_{\rm NL}^{(c)}$ is the breakdown scale for pure capillary waves,
\begin{equation}
\label{eq-k_NL_capillary} k_{\rm NL}^{(c)} \approx
P^{\frac{2}{3}}\sigma^{-1}.
\end{equation}
Inserting expressions (\ref{eq-k_NL_capillary}) and (\ref{eq-k_0})
this gives us the same condition, (\ref{eq-flux}), on the flux! We
see that there is a critical energy flux, $P_c=(g\sigma)^{3/4}$
which can
 be carried by the weak turbulence spectra. The issue of what happens if
$P > P_c$ is of paramount interest. It is clear that in this case there
is a window in $k$ space corresponding roughly to
$\left[k_{\rm NL}^{(c)}, k_{\rm NL}^{(g)}\right]$ where the nonlinearity is
not weak and the dynamics is presumably dominated by fully nonlinear
structures. This situation is illustrated schematically in figure
\ref{fig-window}. It is suggestive that the value of $P_c$, if expressed in
terms of the wind speed, corresponds roughly to the threshold for the
formation of whitecaps on the ocean surface\cite{newell92}.

The phenomenon of intermittency is thought to be associated with
the generation of such strongly nonlinear structures and would
manifest itself in a deviation of the structure functions,
$S_N(\V{r}_1, \ldots \V{r}_{N-1})$, of the wave field from joint
Gaussianity. If one assumes that the statistics are dominated by
whitecaps then one can estimate the scaling behaviour of field
gradients. However, the support of the set of singularities need
not be simple set. There are reasons to expect that whitecaps are
supported on a fractal set of dimension $0\leq D \leq
1$\footnote{The authors would like to thank Prof. V.E. Zakharov
for sharing his observations on this topic.}  although we will
consider fractal sets up to dimension $2$. In this case the
$n^{th}$ moment of the field gradients scales as
\begin{equation}
S_N(\V{r}) \sim (\Delta \theta)^N\left(\frac{r}{L}\right)^{(2-D)},
\end{equation}
where $\Delta \theta$ is a characteristic size of the jump
discontinuities in the derivative and $L$ is the integral scale.
It then follows (see \cite{frisch} sec. 8.5) that,
\begin{equation}
\label{eq-fractalScaling} \frac{S_{2N}(\V{r})}{(S_2(\V{r}))^N}
\sim \left(\frac{r}{L}\right)^{(1-N)(2-D)}.
\end{equation}
For $D<2$ the system deviates from joint Gaussianity.  Such
behaviour is generally thought to be beyond the standard picture
of weak turbulence. Nonetheless, Biven, Nazarenko and Newell
\cite{biven2001,newell2001} have calculated the first correction
to joint Gaussian statistics in the case where the weak turbulence
approximation breaks down. For $N$ even,
\begin{equation}
\label{eq-WTscaling}
\frac{S_{2N}(\V{r})}{(S_2(\V{r}))^N} = 1+\sum_{i=1}^{N/2} C_{N_i}\left(P^{1/3}r^{\alpha-\frac{\gamma}{3}}\right)^{2i-1} + \ldots
\end{equation}
Breakdown occurs for $\gamma>3\alpha$ in which case, the second
term in (\ref{eq-WTscaling}) scales like
$r^{(\alpha-\gamma/3)(N-1)}$ as $r\to 0$. If we wish to attribute
this breakdown to the emergence of whitecap-dominated behaviour,
we observe that it is possible to match the scalings
(\ref{eq-fractalScaling}) and (\ref{eq-WTscaling}) for {\it all N}
if we choose
\begin{equation}
\label{eq-fractalDimension} D = \alpha -\frac{\gamma}{3} +2.
\end{equation}
For gravity waves, $\alpha=1/2$ and $\gamma=3$ so the dimension of the set
of whitecaps would be 3/2. It would be nice if one obtained a value
of $D$ less than one but there are several reasons why this argument is
an oversimplification. In particular, expression (\ref{eq-WTscaling})
represents only the first terms in an infinite  series.
It is highly likely that in
the regime where weak turbulence breaks down, the higher order corrections
which are neglected here actually contribute strongly. Nonetheless it is
a nontrivial fact that this matching can be done consistently for all
values of $N$ simultaneously, even if the actual value of the fractal
dimension obtained here must be considered with caution.

\section{Summary and Conclusion}
Our aim in this article was to show that many of the commonly
considered applications of weak turbulence possess the incomplete
self-similarity property, which can be exploited to obtain core
results using a simple dimensional argument without resorting to
the more complex methods required in general. For such systems,
recent results on the breakdown and range of applicability of weak
turbulence can also be obtained in a simple way. It was found that
dimensional considerations rule out the development of small scale
intermittency in most physically relevant three-wave systems.

We considered the gravity-capillary surface wave system in more
detail and discussed, from the point of view of the dimensional
quantities present, how the validity of the weak turbulence
approximation depends only on the energy flux input at the largest
scales. Even in the case where this flux is large enough to cause
breakdown of the gravity dominated part of the spectrum, we still
have a consistent mechanism to transfer energy to the viscous
scale which consists of two different weakly turbulent regimes
connected by a window of scales where the nonlinearity is strong.
We made some speculative observations about the relationship
between the breakdown of weak turbulence and the emergence of
whitecaps in this window of strong turbulence.

It is appropriate that we finish with some balancing remarks about
situations in which the simple approach outlined here does not
work. Firstly, there are obviously cases of physical interest
which are self-similar. Some important examples are provided by
optical waves of diffraction in nonlinear dielectrics and the
turbulence of waves on Bose-Einstein condensates, both of which
are described by the Nonlinear Schrodinger equation
\cite{dyachenko92}. In these cases, there are two relevant
dimensional parameters.

Secondly, even in the case of incompletely self-similar systems, a
cautionary note should be sounded. The long time behaviour of
these systems is determined by the kinetic equation for the
spectral wave action density,
\begin{equation}
\label{eq-full_KE} \pd{n_k}{t} =
T_2\left[n_k\right]+T_4\left[n_k\right] + \ldots,
\end{equation}
and a nonlinear frequency modulation,
\begin{equation}
\omega_k \to \omega_k + \Omega_2\left[n_k\right] +
\Omega_4\left[n_k\right]+ \ldots.
\end{equation}
For 4-wave systems with an interaction coefficient,
$T_{kk_1k_2k_3}$,
\begin{eqnarray}
 T_2\left[n_k\right] &=& 0,\\
 T_4\left[n_k\right] &=& \int
\left|T_{kk_1k_2k_3}\right|^2n_1n_2n_3\delta(\omega_{01,23})\delta(\k_{01,23})d\k_{123}+
n_k\Im \Omega_4\left[n_k\right],
\end{eqnarray}
with the frequency modulation integrals given by
\begin{eqnarray}
\Omega_2\left[n_k\right] &=& \int T_{kk_1kk_1}n_1\ dk_1, \\
 \Im \Omega_4\left[n_k\right] &=& \int
\left|T_{kk_1k_2k_3}\right|^2(n_1n_3+n_1n_2-n_2n_3)\delta(\omega_{01,23})\delta(\k_{01,23})d\k_{123}.
\end{eqnarray}

Notice that the imaginary part of the frequency modulation enters
the collision integral.  It has been pointed out by Zakharov and
others that if the interaction coefficient is not uniformly
homogeneous in its arguments,
\begin{equation}
T_{kk_1k_2k_3} \sim
(kk_2)^{(1-\chi)\gamma}(k_1k_3)^{\chi\gamma}\hspace{1cm}
0\leq\chi\leq1,
\end{equation}
for example, the frequency correction, $\Im
\Omega_2\left[n_k\right]$ can be divergent at low $\k$, even
though $T$ is still homogeneous of the same degree, $\gamma$.
Luckily, this divergence cancels in the kinetic equation to lowest
order, $T_4\left[n_k\right]$. However, the divergences in the
$\Omega$'s may resurface at higher orders in the full collision
integral, (\ref{eq-full_KE}). It is not clear whether we are
rescued by such cancellations as occur in $T_4\left[n_k\right]$.

If these divergences persist then the dimensional argument applied
in section \ref{sec-breakdown} to estimate the range of validity
and breakdown scales would be not valid and would require
modification to include some non-universal dependence on $\chi$
and a low $\k$ cutoff. Thus wave turbulence may possess a
mechanism for breaking incomplete self-similarity even without the
need to introduce additional dimensional parameters. Whether this
occurs in reality is, at present, an open question.

\section*{Acknowledgements}
We are grateful for financial support from NSF grant 0072803, the
EPSRC and the University of Warwick.

\bibliography{main}
\end{document}